\input{aipcheck}
\documentclass[
    ,final            
  ]  {aipproc}
\layoutstyle{6x9}

\def\etal{\emph{et al.}}
\newcommand{\PRL}[3]{Phys. Rev. Lett., {\bf {#1}}, {#2} ({#3})}

\newcommand{ \be }{\begin{equation}}
\newcommand{ \ee }{\end{equation}}
\newcommand{ \bea }{\begin{eqnarray}}
\newcommand{ \eea }{\end{eqnarray}}
\newcommand{ \la }{\langle}
\newcommand{ \ra }{\rangle}
\newcommand{ \eps }{\varepsilon}

\newcommand{\mean}[1]{\la {#1} \ra}
\def\v2#1{v_2\{#1\}}

\begin{document}

\title{Energy and system size dependence of elliptic flow:
using rapidity gaps to suppress non-flow contribution}


\author{Sergei A. Voloshin}{
  address={Department of Physics and Astronomy, Wayne State
  University, Detroit, 48201}
}

\author{the STAR Collaboration}{}

\begin{abstract}
In this talk I present new STAR results on measurements 
of integrated elliptic flow
at midrapidity in Au+Au and Cu+Cu collisions at 
$\sqrt{s_{NN}}=$200 and 62 GeV energies.
These results have been obtained from azimuthal correlations between
particles in the main STAR TPC and two forward TPCs, and
are to a large extent free from so-called non-flow correlations.
These results along with the previously reported values of
``participant'' eccentricity taking into account eccentricity fluctuations
are used for testing the $v_2/\eps$ scaling, which is found to hold
relatively well. 
\end{abstract}

\maketitle


\section{Introduction}

Anisotropic flow, and, in particular, elliptic flow
plays a very
important role in the understanding of heavy ion collisions at high
energies.
Since the discovery of the in-plane elliptic flow 
at AGS~\cite{e877-flow}, and later
at CERN SPS~\cite{na49-flow-PRL}, the attention to this phenomena has been
continuously increasing. 
The first RHIC measurements~\cite{star-flow1} by the STAR Collaboration
made it one of the most important observable at 
RHIC~\cite{voloshin-qm2002}. 
The large elliptic flow, along with its
mass dependence at low transverse momenta~\cite{star-flow2,v2pt-mass}
are often used as a strong argument for early thermalization of the
system;
the observed~\cite{star-flow-cqs} constituent quark
scaling~\cite{voloshin-qm2002,voloshin-molnar} 
of elliptic flow strongly suggests that the system spends significant
time in the deconfined state.

In order to understand deeper the physics of the elliptic flow and the
processes governing the evolution of the system, one has to measure
and understand elliptic flow dependence on the system size and
collision energy. 
Having in mind also the dependence on the centrality of the
collision makes the problem indeed 'multi-dimensional'.  
In~\cite{voloshin-poskanzer-PLB} the authors suggested that the
elliptic flow follows a simple scaling in the initial system
eccentricity and the particle density in the transverse plane, 
$v_2/\eps \propto 1/S~dN_{ch}/dy$
(where $S$ is the area of the overlap region of two nuclei). 
If true, all the data taken at different energies, colliding different
nuclei, and at different centralities should collapse on the same universal
curve. 
Indeed the available data are consistent with such a
scaling~\cite{voloshin-qm2002,na49-flow-PRC}. 
Unfortunately, the systematic uncertainties in the results 
are too large to conclude on how well the scaling holds. 
The main difficulties are the evaluation/elimination  of the
so-called non-flow correlations (azimuthal correlations not related to
the reaction plane orientation), and effects of flow fluctuations, 
along with uncertainties in the calculation of the initial
system eccentricity needed for such a plot. 

The non-flow contribution is usually suppressed when one uses
multi-particle correlations to measure anisotropic 
flow~\cite{olli4,star-flow-PRC}. Unfortunately, 
multi-particle correlations are sensitive to flow fluctuations in very
different way, thus leaving the problem of disentangling of non-flow
contribution and flow fluctuations unresolved.
An alternative way to suppress non-flow contributions, pursued in this
analysis, is the correlation of
particles with large rapidity gaps, the main focus of this analysis.
In particular we present the results for anisotropic flow at
midrapidity measured in the STAR main TPC ($-0.9<\eta<0.9$) from
correlations with particles in the two Forward TPCs ($2.9<|\eta|<3.9$).
Detailed study of these correlations indicate that in this case the
non-flow contribution is suppressed up to about factor of 4-5 compared to
the case of particle correlation only in the main TPC region. 

The results presented in this talk are based on the analysis (after
all event quality cuts) of 5.9 M Au+Au 200 GeV, 7 M Au+Au  62 GeV, 3.5 M
Cu+Cu 200 GeV, and 19 M Cu+Cu  62 GeV   Minimum Bias events.
Centrality of the collision was determined in accordance with the
so-called Reference Multiplicity - the multiplicity of primary tracks
in $|\eta|<0.5$ region. For Au+Au collisions the centrality
determination was done directly from data taking into account 
 known vertex reconstruction efficiency. For Cu+Cu collision, where
vertex reconstruction efficiency is still under study, the centrality
was defined based on Glauber Monte-Carlo calculations (with additional
simulation of charged particle production per nucleon participant)
matching the experimental distribution at high multiplicities, where
the vertex reconstruction efficiency is close to 100\%.

Fig.~1 and Fig.~2 presents the results for elliptic flow of charged
particles in the pseudorapidity window $|\eta|<0.9$, 
for all four systems studied in this analysis. 
Charged particles are selected from $0.15<p_t<2.0$~GeV region; the low
transverse momentum cut is due to TPC acceptance. 
In black, noted as $v_2\{2\}$, 
are shown the results obtained from two particle azimuthal correlations with 
both particle from the main TPC region. 
In blue, noted as $v_2\{\rm{FTPC}\}$,  are the results obtained from
correlation with large rapidity gap (correlating particles in the main
and Forward TPCs regions). The larger values of  $v_2\{2\}$ compared
to $v_2\{\rm{FTPC}\}$ are attributed to the non-flow contribution.
The relative contribution of non-flow in Cu+Cu collisions is
significantly larger compared to that in Au+Au collisions due to smaller values
of flow itself. 
Cu+Cu results exhibit also weaker centrality dependence, which can be
explained by weaker correlations of Reference Multiplicity and impact
parameter.

\begin{figure}
  \includegraphics[width=.48\textwidth]{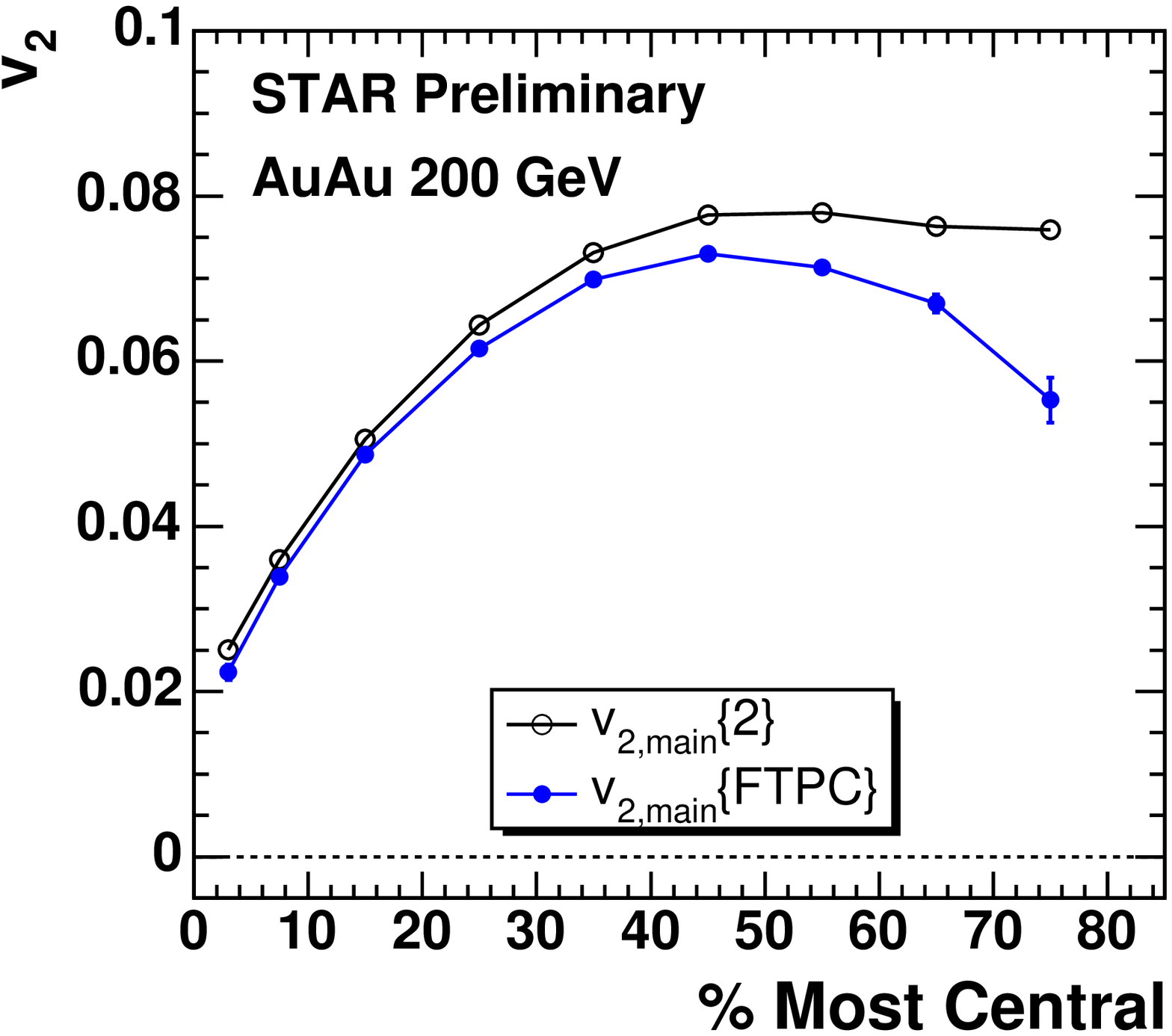}
  \includegraphics[width=.48\textwidth]{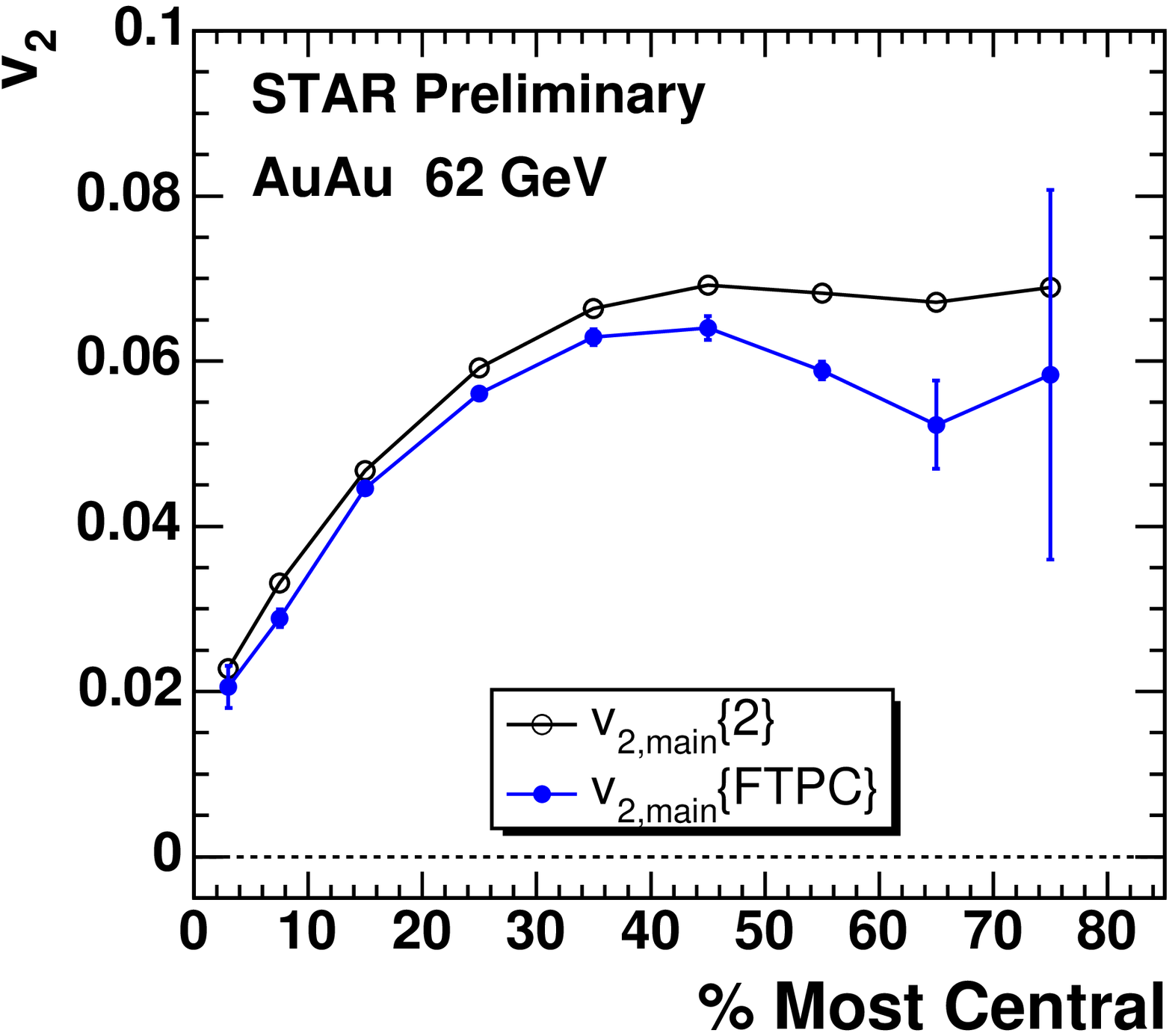}
  \caption{Elliptic flow in Au+Au collisions as function of centrality
  at $\sqrt{s_{NN}}=$200 and 62 GeV.}
\end{figure}
\begin{figure}
  \includegraphics[width=.48\textwidth]{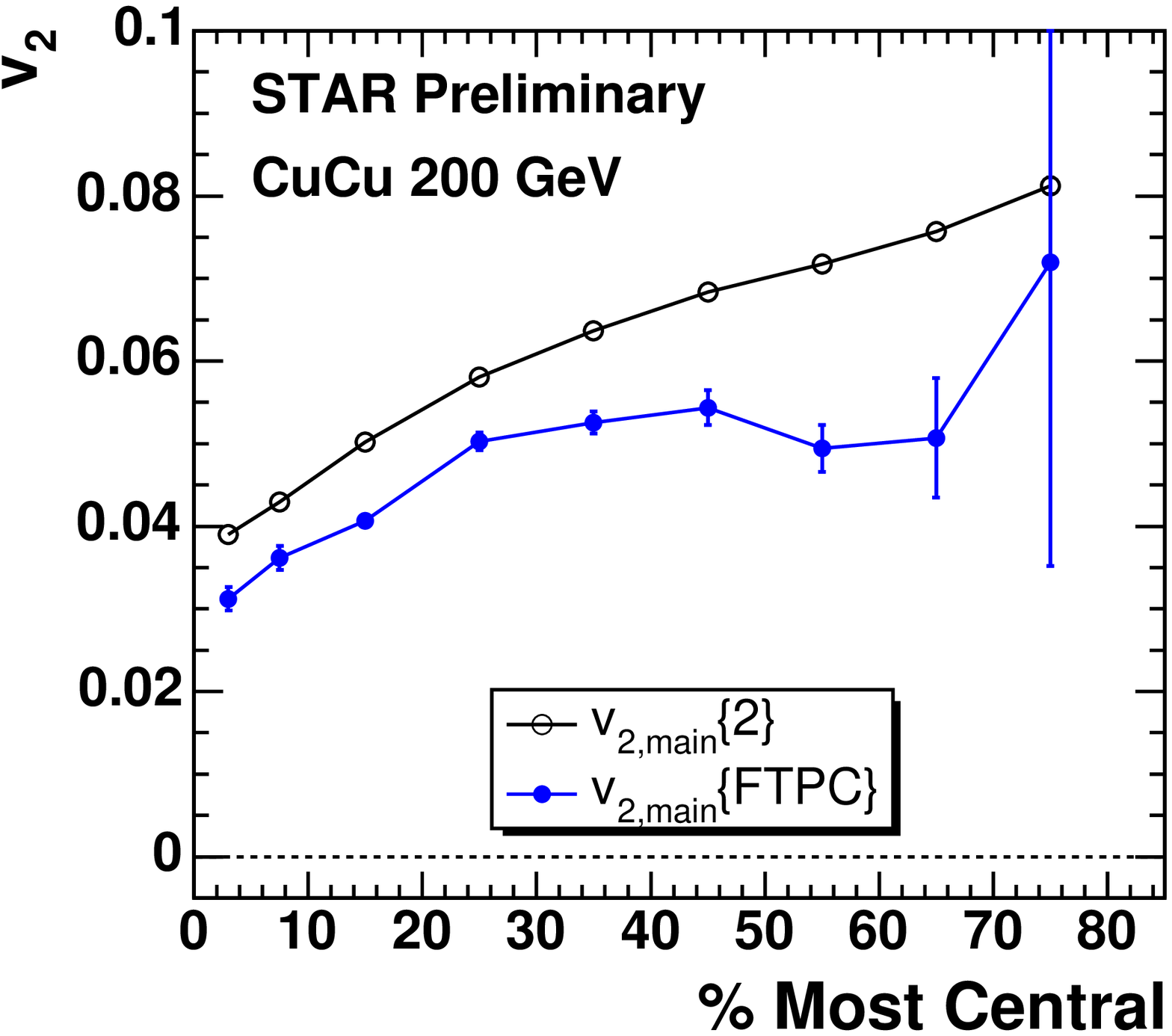}
  \includegraphics[width=.48\textwidth]{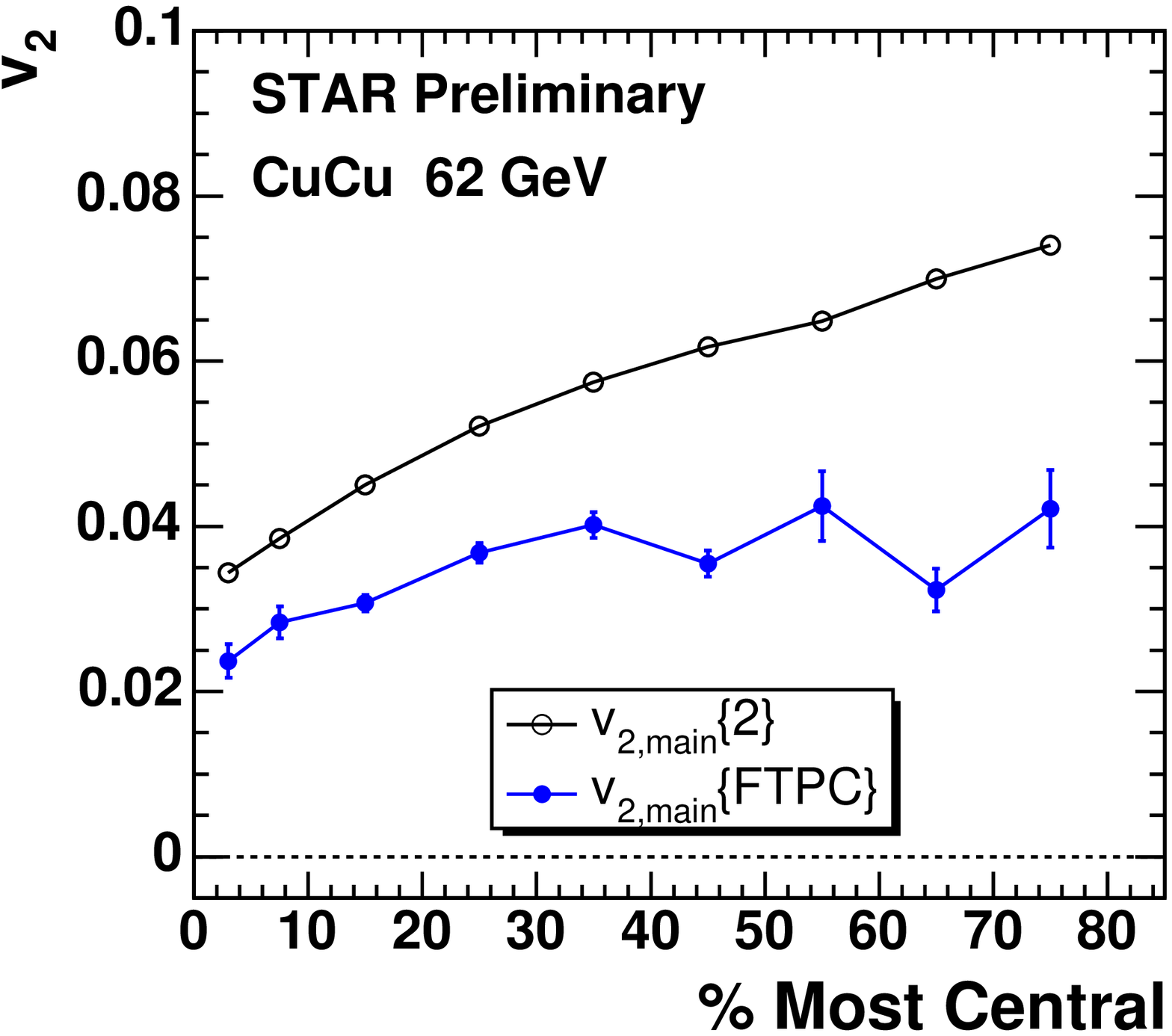}
  \caption{The same as Fig.~1 for Cu+Cu collisions.}
\end{figure}

Besides suppression of non-flow contribution, an adequate treatment of
flow fluctuations is essential for understanding the flow dependence on
centrality of the collision and the size of colliding nuclei. 
Usually it is assumed that elliptic flow
closely follows initial eccentricity of the system, 
$\eps=\mean{y^2-x^2}/\mean{y^2+x^2}$. 
Under this assumption one can estimate flow fluctuations by calculation
of eccentricity fluctuations~\cite{star-flow-PRC,raimond-mike}, 
e.g. in Glauber Monte-Carlo model. It was
argued~\cite{manly-qm2005,voloshin:LaJolla} that the so-called
``participant'' eccentricity should be used in this calculations.
The corresponding results for the four systems studied in this analysis
have been presented
in~\cite{voloshin:LaJolla}, and are used in the current analysis.

The final plots of this presentation, Fig.~3, show how new
results fit to the $v_2/\eps$ scaling. 
Two plots are made under different assumptions for how strongly flow
fluctuations in the main TPC region, $|\eta|<0.5$, are correlated with
flow fluctuations in the Forward TPC regions $2.9<|\eta|3.9$. 
If the fluctuations are totally independent one has to use
$\mean{\eps_{part}}$ for such a plot. 
This is shown on the left panel.
If flow fluctuate coherently in the main TPC region and in Forward TPC
regions, one should use 
$\eps_{part}\{2\}=\sqrt{\mean{\eps_{part}^2}}$ 
for the scaling plot, shown in the right panel.
The scaling in latter case is somewhat better. 
More details on how flow results obtained in different
ways should scale with eccentricity can be found in\cite{voloshin:LaJolla}. 
For the references to other data presented in Fig.~3 see
in~\cite{na49-flow-PRC}. 

Note that for the  $v_2/\eps$ scaling plots the results from Figures 1
and 2 have been
rescaled/extrapolated to a full transverse momentum coverage using
blast wave fits to spectra; rapidity density has been obtained from
pseudorapidity density using scaling factors based on HIJING and RQMD
calculations. 
The presented (ideal) hydrodynamic predictions are based on
calculations~\cite{hydro}. The curves shown are obtained from hydro
results made for fixed impact parameter ($b=7$~fm) and different
particle densities (collision energies). 
Note that hydro results do not scale perfectly in
this plot and in general exhibit somewhat flatter centrality
dependence at each collision energy.

\begin{figure}
  \includegraphics[width=.58\textwidth]{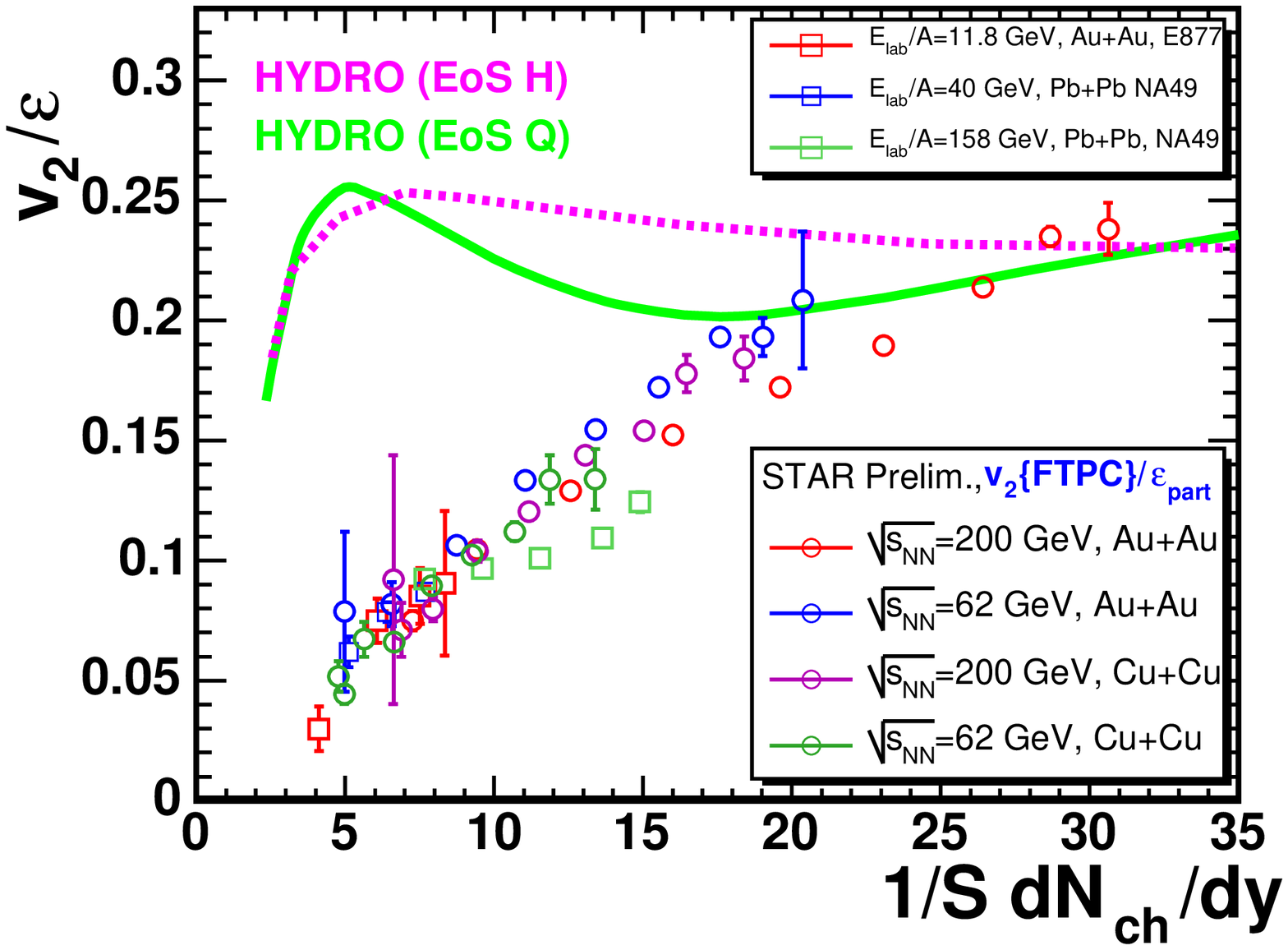}
	\hspace*{-0.8cm}
  \includegraphics[width=.58\textwidth]{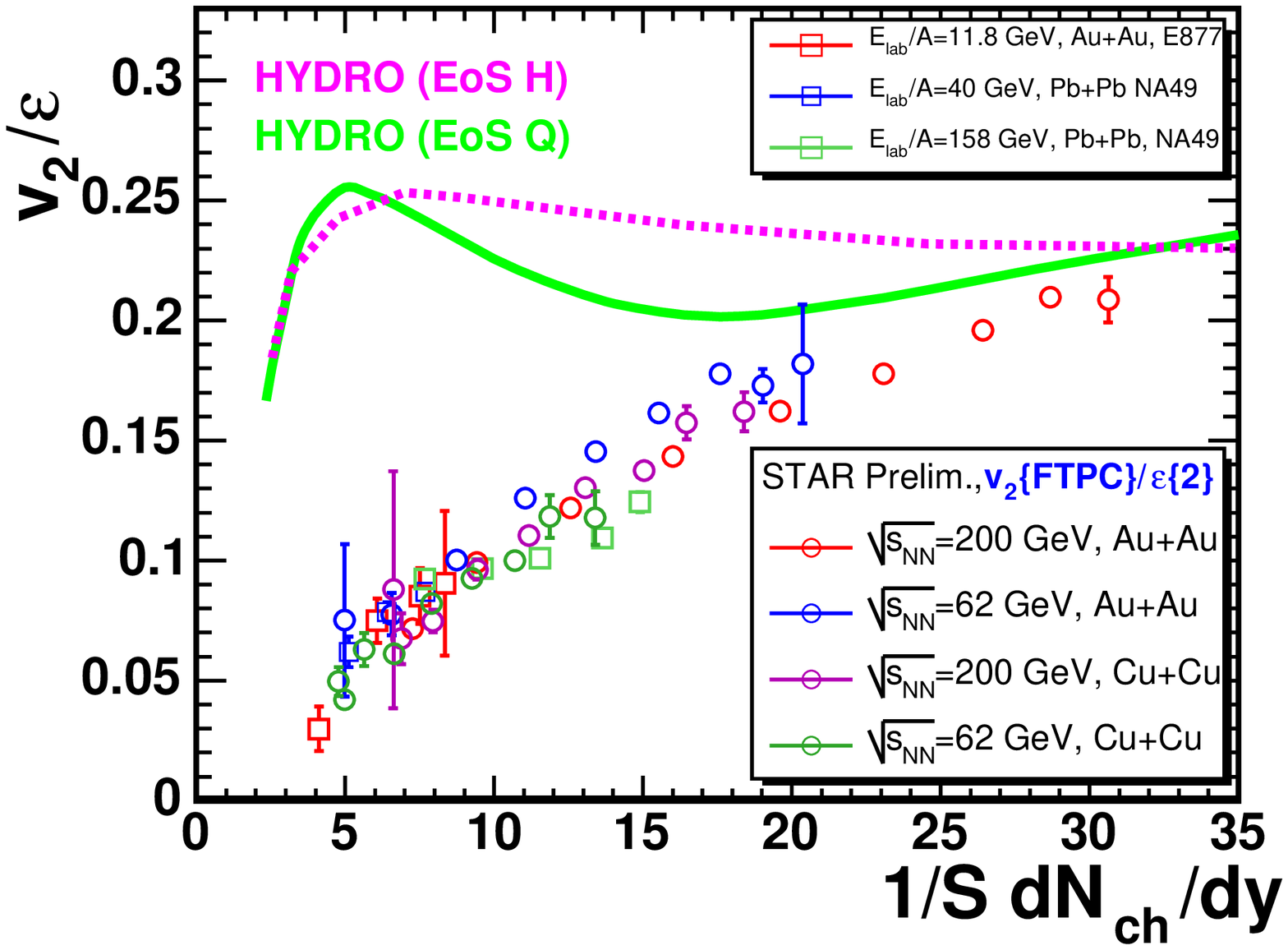}
  \caption{$v_2/\eps$ scaling plot using $v_2\{\rm{FTPC}\}$ results and
  rescaled with $\mean{\eps_{part}}$ (left panel) and
  $\eps_{part}\{2\}$ (right panel). }
\end{figure}

\begin{theacknowledgments}
I thank the organizers for a very interesting and stimulating
conference and giving me the opportunity to present our results.
Financial support provided by US Department of Energy Grant No. DE-FG02-92ER40713.
\end{theacknowledgments}

\end{document}